\documentstyle[psfig]{lamuphys}
\makeatletter
\let\chapter\hid@chapter
\makeatother
\begin{document}
\pagenumbering{arabic}
\title{The N~V/C~IV ratio in high redshift radio galaxies}

\author{R.\ A.\ E.\ Fosbury\inst{1}, J.\ Vernet\inst{2}
M.\ Villar-Mart\'\i n\inst{3}\\
 M.\ H.\ Cohen\inst{4}, A.\ Cimatti\inst{5}, S.\ di Serego Alighieri\inst{5}}

\institute{ST-ECF, Garching bei M\"unchen, Germany
\and
ESO, Garching bei M\"nchen, Germany
\and 
Institute d'Astrophysique, Paris, France 
\and
California Institute of Technology, Pasadena, USA
\and
Osservatorio Astrofisico di Arcetri, Firenze, Italy}

\authorrunning{R A E Fosbury et al.}

\maketitle

\begin{abstract}
Deep Keck spectropolarimetry has been obtained of a sample of powerful radio
galaxies with $z\sim2.5\pm1$. In addition to a clear 2200\AA\ dust
scattering/extinction signature and an inverse correlation between the degree
of continuum polarization and the Lyman-$\alpha$\ emission line strength
(relative to C~IV), we find an intriguing positive correlation between the
N~V/C~IV ratio and the polarization. The line ratio, which varies by almost a
dex amongst objects having a very similar ionization state, is most likely to
indicate an abundance ratio. We speculate that the correlation with
polarization represents a connection between nitrogen enhancement and dust
production following a major starburst event connected with the birth of the
quasar. 
\end{abstract}
\section{Introduction}
Our current belief is that the hosts of powerful radio sources in the distant
Universe are destined to become the giant ellipticals of today: the most
massive galactic systems we know. While some may have commenced their formation
at very high redshift, it is clear that the process of assembly is very active
at $z\sim2.5$. This corresponds to the epoch when quasars appear to have had
their maximum space density (Dunlop 1994). Indeed, a major goal of our
programme is to understand the causal relationship between the formation of the
massive black hole and that of the galaxy within which it resides. Our approach
has been to identify, catalogue and measure the components which contribute to
the observed spectral energy distribution of active galaxies at different
redshifts. The realisation that reprocessed AGN radiation makes a very
significant contribution to the blue and ultraviolet spectrum indicated how
difficult it is to measure the stellar population in the galaxy using these
regions of the restframe spectrum. The energetic radiation, and perhaps
mechanical energy, from the hidden quasar does however illuminate and excite
the host's ISM and --- provided we understand the physics of the ionization
processes --- gives us a unique opportunity to analyse this important
constituent of the galaxy.

Our approach of observing radio galaxies rather than the hosts of objects
classified as quasars relies on the now well-justified assumption that these
are similar objects viewed from different directions. The presence of very
opaque material in the nuclear regions of the galaxy results in a `natural
coronograph' which greatly facilitates the study of the underlying stellar
galaxy and its ISM. The use of spectropolarimetry to study the scattered
radiation from the hidden quasar in radio galaxies has shown that the
reprocessed AGN light is a major contributor to the `alignment effect' (eg.
Tadhunter et al. 1988; Cimatti et al. 1993; di Serego Alighieri et al.
1993, 1994; Jannuzi et al. 1995).  The combination of spectroscopic and
polarimetric measurements is a powerful discriminant between such reflected AGN
radiation and stellar and nebular continua arising on a galactic scale.

\section{The sample}

We have selected powerful RG with $z\sim2.5$\  which allow us to study in
visible light the strong UV emission lines from Ly-$\alpha$\ to CIII], the UV
continuum, the resonance absorption lines from the ISM and from OB stars and
the 2200\AA\ dust feature and also to straddle the 4000\AA\ break in the
1--2$\mu$m band.

\begin{table}
\begin{center}
\caption{The sample sources and their redshifts. The last four sources are 
taken from similar data in the literature and plotted with a different symbol 
in the figures.}

\begin{tabular}{lr}
OBJECT   &  REDSHIFT \\
4C+03.24&3.570 \\
MRC 0943-242&2.9 \\
USS 0828+193&2.572 \\
4C+23.56a&2.482 \\
B3 0731+438&2.429 \\
4C-00.54&2.366 \\
4C+48.48&2.343 \\
TXS 0211-122&2.338 \\
4C+40.36&2.265 \\
4C+41.17&3.798 \\
FSC10214+4724&2.282 \\
MRC 2025-218&2.63 \\
3C 256&1.824 \\
\end{tabular}
\end{center}
\end{table}

Our principal sample constists of nine objects with $2.3 < z < 3.6$\ and this
is suplemented by four sources from the literature having similar quality data
but extending the redshift range to $1.8 < z < 3.8$. More data are currently
being obtained on sources with $z>3$.

\section{Observations}

The Keck spectropolarimetric observations for the first two sources in
the programme are described in Cimatti et al. (1998). The observational
procedure for the remaining sources was similar and resulted in total exposure
times ranging from 4--8~hr per source.

The HST WFPC~2 and NICMOS images, where available, were taken from the public
archive at the ST-ECF and from the McCarthy et al. program (ID 7498). These are
being compared with the one dimensional line and continuum extensions measured
from the spectra.

\section{Results}

The spectropolarimetry gives us line and continuum polarization measurements
with good s/n, ranging up to 20 or so in the continuum longward of Ly-$\alpha$\
for the highly polarized sources. The measured fractional polarizations range
from $\le 1$\% to nearly 20\%. A source with an intermediate polarization 
in our sample,
USS~0828+193, is illustrated in Figure~1. The narrow emission lines have lower
polarization than the continuum. In the average spectrum of our sources,
the broad lines can be seen clearly in the
polarized flux with equivalent widths similar to those seen in quasars. 
The position angle of the electric vector is perpendicular to the
UV extensions in the images which usually, but not invariably, corresponds to
the radio axis. The continuum polarizations generally increase slightly to
shorter wavelengths.

\begin{figure}[!b]
\centerline{
\psfig{figure=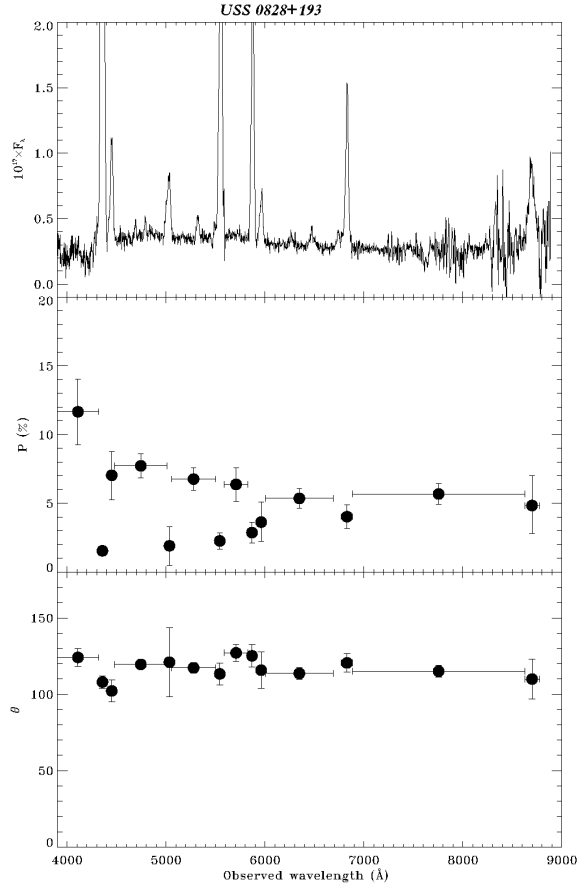,width=8cm,angle=0}
}
\caption{Spectropolarimetry of USS~0828+193 covering restframe wavelengths from 
shortward of Ly-$\alpha$\ to 2450\AA. This object shows moderate continuum polarization
and a rich emission line spectrum. It also shows the characteristic continuum
peak just longward of Ly-$\alpha$\ and a dip near a rest wavelength of 2200\AA.
The three panels contain: top -- total flux, middle 
-- fractional polarization and bottom -- PA of the E-vector. The horizontal bars 
delineate the wavebands within which $P$\ and $\theta$\ are calculated. The error
bars are 1$\sigma$.}
\end{figure}

The first new result concerns the continuum spectrum of the objects. In the
restframe, the spectra have a rather uniform and characteristic shape with a
peak (in $f_{\lambda}$) just longward of Ly-$\alpha$\ and a shallow minimum
near 2200\AA. This is discussed further in the contribution by Vernet et al.
(1999) in these proceedings and can be explained, with remarkable precision, by
reflecting an average quasar spectrum in a scattering `atmosphere' containing
standard Galactic dust. The dust grains scatter and absorb in such a way that
the reflected light is maximised when $\tau_{scat}\sim\tau_{ext}\sim 1$. Such a
process results in approximately `grey' scattering but the difference between
$q_{scat}$\ and $q_{ext}$\ results in the imposition a 2200\AA\ signature onto
the spectrum. This is clearly distinct from the Fe~II emission feature just
longward of this wavelength which is a characteristic of quasar spectra. The
precision of the fit of this simple one parameter model (the fraction of the
quasar light which is scattered) to the observed spectra leaves little room for
other sources of continuum such as a young stellar population which Vernet et
al. (1999 and in prep.) show must be less than about 20\% of the reflected light. 

The second result is a clear anticorrelation between the strength of the
Ly-$\alpha$\ emission line (relative to C~IV) and the continuum polarization,
illustrated in Figure~2. While it is likely that this is a consequence of the
resonant destruction of Ly-$\alpha$\ photons by the dust grains which scatter
the polarized light we observe, the detailed explanation will involve an
understanding of the geometric details and will not be discussed further here.

\begin{figure}[!b]
\centerline{
\psfig{figure=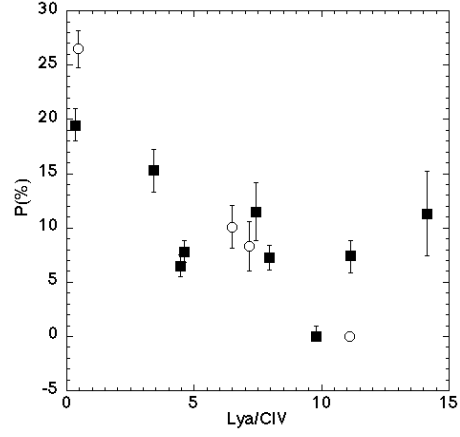,width=6cm,angle=0}
}
\caption{The relationship between the continuum polarization, measured just 
longward of Ly-$\alpha$--N~V, and the Ly-$\alpha$/C~IV ratio. In this and in 
Figure~3, the objects from our sample are plotted as filled squares. The error 
bars are 1$\sigma$ uncertainties in the polarization. The statistical 
uncertainties in the line ratios are too small to plot.}
\end{figure}

The third result, and the principal subject of this talk, concerns the
behaviour of the N~V resonance line at 1240\AA. While both the continuum
colours and the relative strengths of the strong emission lines of C~IV, He~II
and C~III] show very little variation amongst the objects in our sample, there
is --- in addition to the Ly-$\alpha$\ effect discussed above --- a large
variation in the ratio of N~V to C~IV. This variation is correlated with the
continuum polarization in the sense that the strongly polarized sources have a
large N~V/C~IV ratio as shown in Figure~3. Such a large range in this ratio, in
objects with such constant C~III], C~IV and He~II lines, cannot readily be
explained as an ionization effect and must indicate a genuine abundance effect.

\begin{figure}[!b]
\centerline{
\psfig{figure=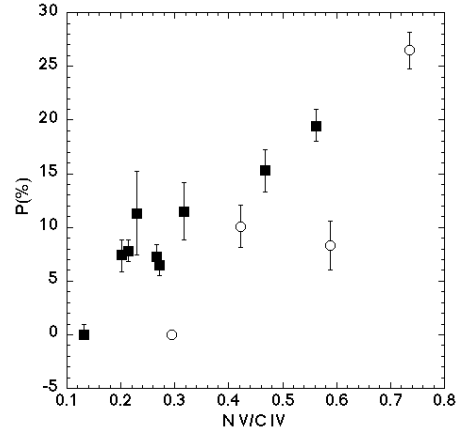,width=6cm,angle=0}
}
\caption{The relationship between the continuum polarization and the N~V/C~IV 
ratio.}
\end{figure}

\section{Discussion}

Although the continuum and emission line regions in the radio galaxies are well
resolved (several arcseconds $\sim$ 10's of kpc), this behaviour of N~V/C~IV is
reminiscent of the behaviour in high redshift QSO broad line regions discussed
by Hamman \&\ Ferland (1993). Indeed, our radio galaxies form a parallel
sequence to the QSOs in their N~V/C~IV -- N~V/He~II diagram. They explain the 
distribution of BLR ratios in this diagram as variations in the metallicity of
the gas resulting from rapid chemical evolution in the massive host galaxies
involving an IMF biassed to more massive stars than characteristic of our
Galactic disk. Our result shows that the abundance variations in these massive
systems are not restricted to the small amount of gas associated with the AGN
BLR but pervade a significant fraction of the ionized ISM.

If the correlation of N~V/C~IV with continuum polarization cannot be
explained as a subtle effect of orientation or geometrical configuration,
then we may be witnessing a direct relationship between dust and primary
nitrogen production in a population of massive stars. Such evolution would occur
on the very rapid timescale of tens of millions of years and the variations
we see in our sample may represent this `micro-evolution'. It is the stage at
which the process is illuminated by the AGN --- presumably fed by the same
event that triggers the star formation --- which determines the global state
of the ISM as we observe it. Thus, rather than observing directly the light
from  a young stellar population which is hidden by the glare of the quasar,
we can observe the chemical effects of the starburst through the properties
of the ISM. The ability to measure emission lines in the resframe optical
spectrum with the new large telescope infrared spectrographs will be crucial
in this analysis.

\noindent
{\bf Acknowledgments} We thank Francesca Matteucci, Jay Gallagher and Alvio Renzini for
discussions. RAEF is affiliated to the Astrophysics Division of the  Space Science
Department, European Space Agency.

%
%

\end{document}